# Specifics of ITO properties deposited on cerium-doped glass for space-grade solar cells

Danil D. Gren[1], Lev O. Luchnikov[1], Dmitri Yu. Dorofeev[1], Prokhor A. Alekseev[2,3], Ildar R. Sayarov[4], Alexey R. Tameev[4], Mikhail S. Dunaevskiy[2,3], Vladislav Kalinichenko[2], Vladimir Ivanov[2], Danila S. Saranin[1,2*] and Eugene I. Terukov[2]

[1]National University of Science and Technology MISIS Moscow, Russia

[2]ITMO University, 197101, St. Petersburg, Russia

[3]Ioffe Institute, 194021, St. Petersburg, Russia

[4]Frumkin Institute of Physical Chemistry and Electrochemistry, Russian Academy of Sciences,

Corresponding author*:

Dr. Danila S. Saranin saranin.ds@misis.ru

## Abstract

Ce-doped glass is a well-established solution for ultraviolet and ionizing radiation shielding of solar cells in space. Traditionally, Ce-glass protected Si or III-V based devices as an overlaying cap. However, for emerging photovoltaics such as halide perovskites, thin Ce-glass coated with transparent conductive layers could serve as a lightweight carrier with an electrode. While indium-tin oxide (ITO) is widely used in solar cells for charge collection, its optical, structural, and electrical properties depend on the substrate quality. In this work, we demonstrated significant differences in properties of ITO deposited on Ce-glass (100 micron thick) compared to standard soda lime glass. ITO on Ce-glass exhibited pronounced compressive strain because of higher oxygen vacancy concentrations, reduced transparency and charge carrier concentration (~$10^{19}$ cm$^{-3}$) resulting from altered stoichiometry. Electrical analysis showed increased Hall mobility (66 cm²/V·s) but decreased conductivity due to excess tin incorporation. These specific ITO features originated from inelastic collisions with the substrate during deposition. Variations in wettability and surface potential underscore substrate-induced differences critical for developing optimized ITO coatings for space-grade photovoltaics.

The rapid advancement in thin-film technology of halide perovskite photovoltaics (**HP PVs**) opens various pathways for its applications[1]. Strong optical absorption and reduced recombination dynamics in microcrystalline HP absorbers allows to reach competitive power conversion efficiency (**PCE**) at standard terrestrial conditions (AM 1.5 G) close to 27%[2]. Various investigations on space applications of perovskite solar cells report the unique radiation resistant properties[3–5] and defect healing effects even under harsh ionizing radiation exposure[6,7]. Real exploitation of solar cells in space conditions requires selection of robust materials for the full device construction. In contrast to wafer-based solar cells like III-V and Si, single junction HP devices should be fabricated on the optically transparent substrates which allow penetration of the light and hold the multilayer device stack in the surface[8]. Exploitation of energy systems in the satellites requires a reduced mass of the devices, thus, the substrate for PSCs should be as thin as possible. Plastics (polyethylene terephthalate (PET) or naphthalate (PEN), typically 100 – 250 microns thick) are efficiently used at terrestrial conditions, but harsh thermal cycling (up to 150°C) at high Earth orbits could be challenging due to bending stress[9]. Standard soda lime glass in thick configuration also couldn't be an option because of darkening effects under radiation and ultraviolet (**UV**) light. Cerium-doped (Ce) glass has emerged as a promising cover material for space photovoltaics, owing to its superior resistance to solarization, UV-induced darkening, and ionizing radiation[10–14]. The incorporation of cerium ions into the glass matrix enhances its ability to absorb and neutralize high-energy photons and charged particles, thereby protecting underlying device layers from radiation-induced damage[15,16]. Traditionally, Ce glass has been employed as a protective cover for silicon and III-V solar cells in satellite technology, but its integration as an active substrate in perovskite photovoltaics remains largely unexplored. The standard architectures of perovskite solar cells and modules have a transparent conductive coating, which acts as charge selective contact. Indium tin oxide (ITO) is the industry standard for transparent conductive electrodes in optoelectronic devices, including perovskite solar cells, due to its exceptional combination of high optical transparency in the visible spectrum and low sheet resistance[17]. ITO's compatibility with various deposition techniques, mechanical flexibility, and chemical stability further reinforce its widespread adoption. Combining the shielding properties of Ce-glass with the conductive functionality of ITO presents a compelling strategy for next-generation perovskite solar cells intended for space and other demanding environments. However, the quality of ITO electrodes can be compromised by the type of substrates for the deposition. Up to date, the specifics of the ITO properties deposited on Ce-glass remain unexplored, while this could be an effective solution with dual functionality.

In this study, we investigated the structural, optical, and electrical properties of ITO films deposited on cerium-doped glass substrates with reactive ion-beam sputtering technique. The comparison of ITO films on standard soda lime glass revealed appearance of morphological non-uniformities, stoichiometry changes, parasitic absorbance and impact to the transport properties (Hall mobility, etc.). Surface properties were estimated via atomic force based adhesion measurements, and Kelvin probe surface potential mapping. Obtained results were deeply analyzed and discussed.

In this work, we used a common soda lime glass (**SLG**) with a thickness of 1.1 mm and Ce-based glass (**CG**) of 100 um. The deposition of ITO was done through reactive ion-beam sputtering with post annealing in ambient conditions. Briefly, the sputtering process was done using In/Sn (90%/10%) alloy target at 3000 V and 130 mA of current flow at the ion source. General schematics for ion-beam sputtering process presented in **fig.1(a)**. The base pressure was $1 \times 10^{-2}$ Pa, and the working gas was Ar (99.998%). $O_2$ oxidation gas was injected into the deposition chamber at 0.06 liter per hour flow. The surface of the target was pre-cleaned with an ion-beam for 10 minutes prior deposition on the substrates. Samples were mounted on the movable substrate carrier, the calculated rate of the deposition was 0.4 nm/min. After deposition, the samples were annealed at 400 °C for 1 hour. For the estimation of the thin-film ITO properties, we deposited 130 nm thick films on the substrates of both configurations. A detailed description of experimental section, methods and characterization presented in Electronic Supplementary Information (**ESI**). To simplify the titling of the sample configuration in the investigation, we will use the abbreviation "**CG/ITO**" for the coating deposited on the Ce-glass and "**SLG/ITO**" for the samples fabricated on soda lime glass.

We used atomic force microscopy (AFM) to estimate the morphology of ITO films deposited on SLG and CG, respectively. Representative topography maps for both samples presented in the **fig.1(b),(c)**. SLG/ITO surface displays a nanocrystalline polymorphic structure with crystallites of approximately ~50 nm. In contrast, the CG/ITO film doesn't exhibit a periodic grain structure. The surface of ITO on the Ce-glass had inhomogeneities with 40 nm in diameter and 2 nm in depth. Also, we observed rounded features with heights up to 6 nm distributed on the surface of the coating. Phase analysis of the coatings was made via–X-ray diffraction (XRD) spectroscopy. The measurements were carried out using a diffractometer in parallel beam geometry. A copper anode X-ray tube (CuKα radiation) served as the source of X-ray radiation. **Fig. 1(d),(e)** show the phase composition analysis for "as-deposited" ITO film with an amorphous structure, which undergoes crystallization into the cubic $In_2O_3$ phase [PDF2:00-006-0416] upon annealing. Rietveld refinement of the XRD patterns indicated a preferential (222) orientation for the CG/ITO sample, with a texture coefficient of 0.49, whereas the SLG/ITO sample exhibits no preferential orientation (texture coefficient of 1.00). For both samples, the diffraction peaks are shifted to higher angles relative to stoichiometric $In_2O_3$. Specifically, the (222) reflection for SLG/ITO was shifted by 0.012 Å, while for CG/ITO the shift is 0.019 Å. The corresponding lattice parameter "a" was calculated as 10.005 Å for SLG/ITO and 10.055 Å for CG/ITO, both significantly reduced compared to the reference value of 10.118 Å for bulk $In_2O_3$. This reduction in lattice parameter suggests the presence of compressive lattice strain so called "elastic effect"[18], likely associated with a high concentration of oxygen vacancies within the crystalline lattice[19].

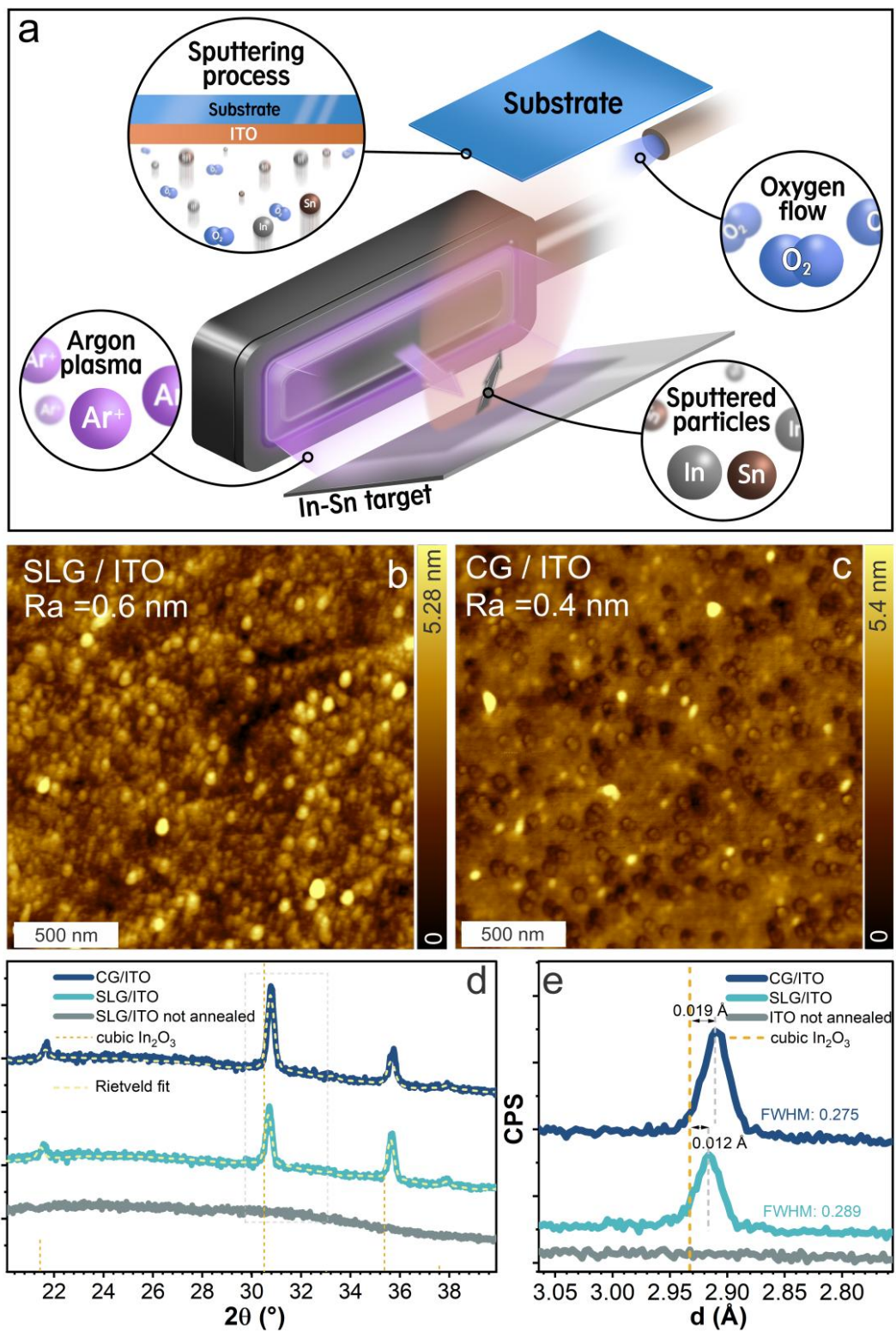

Figure 1 – Scheme of IBS ITO deposition (a), Surface morphology of SLG/ITO (b), CG/ITO (c), XRD pattern of ITO (d) and (e)

We studied optical properties of ITO coating deposited on different types of glass through transmittance (T) measurements and Tauc plot calculations. Photo-images of the fabricated samples. The enhanced absorption for the CG/ITO samples could be visually estimated as presented in photo-images of the **fig.2(a),(b)**. Strong absorption of the Ce glass in the UV region shifted to a rise of transmittance spectra from 350 nm for SLG to 370 nm. The UV absorption in cerium oxide primarily arises from its electronic structure and the redox chemistry of cerium. $CeO_2$ exhibits strong absorption in the UV region due to presence oxygen vacancies and $Ce^{3+}$ states[19]. UV irradiation leads to the generation of oxygen vacancies, where $Ce^{4+}$ ions are partially reduced to $Ce^{3+}$[19]. These defects introduce defect states within the band-gap, which enhanced the absorption of photons in the UV and visible range. 100 micron thick Ce-glass had a transmittance of 91%, with a plateau up to 1000 nm. The T-spectra for SLG showed a typical value of ~93% in the visible range with a reduction to 87% in the near infra-red region. Maximum transmittance value of 89% for SLG/ITO was measured at 560 nm. In the longer wavelength of ~800 nm, T for SLG/ITO reduced to 78% remains almost constant till 1000 nm. The absorption of CG/ITO was increased in the short wavelength, and transmittance doesn't exceed 79% over entire range of the wavelengths used in measurements.

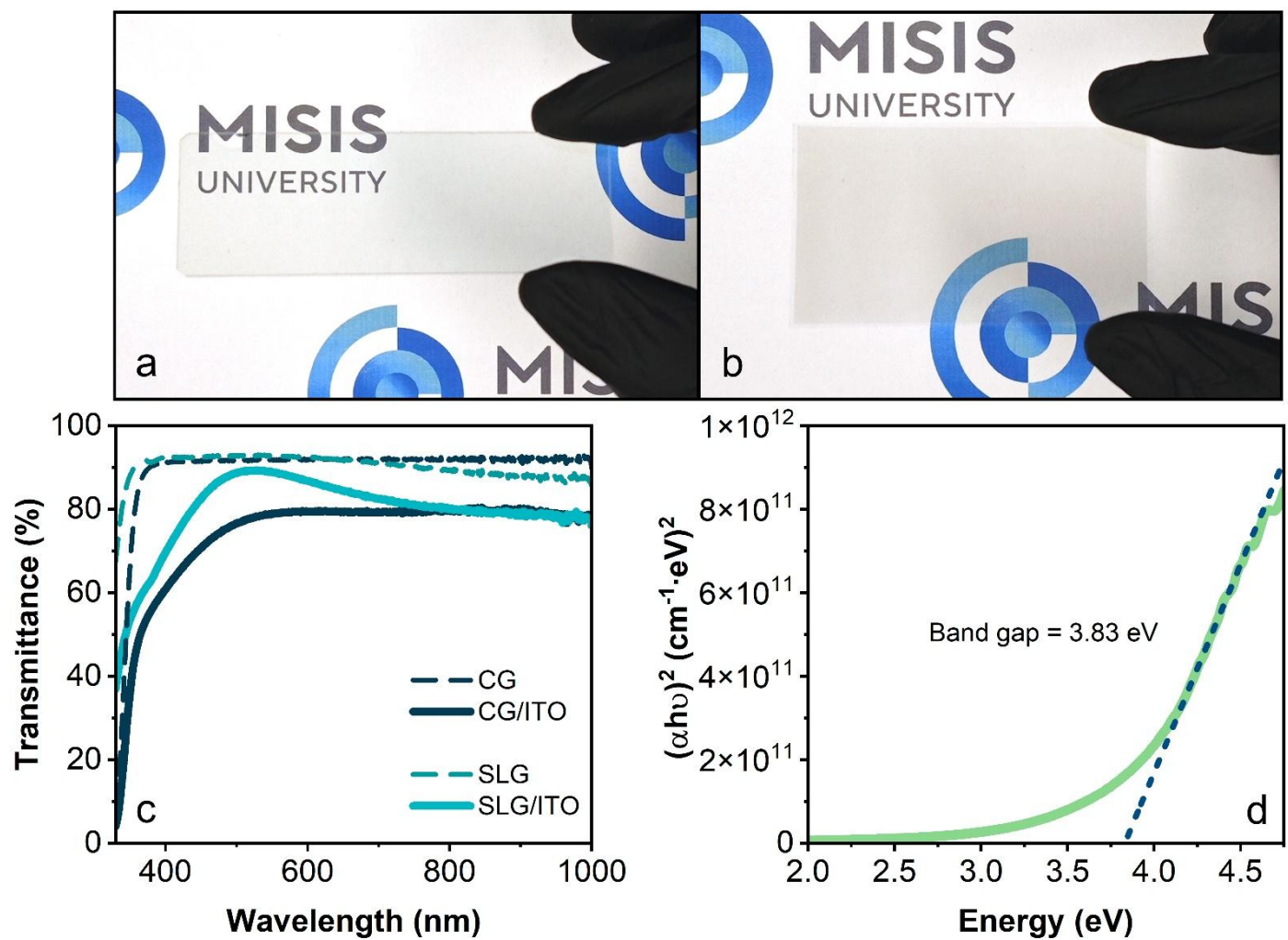

Figure 2 – Photo-images of the ITO coatings deposited on SLG (a) and CG(b) substrates; transmittance spectra in VIS and NIR ranges (c); Tauc plot approximation for ITO coating deposited on quartz (d)

ITO is a wide band-gap semiconductor, and gathering precise values from the samples on UV-absorbing substrates is complex. We used quartz glass with ion-beam deposited ITO to extract band-gap ($E_g$) via Tauc method[20] (**eq.1** in **ESI**). The calculated value was 3.82 eV, which is in agreement with the published data in the literature[21–23] for the reactively fabricated ITO films.

Electric properties were estimated via Hall mobility($\mu_{Hall}$) measurements and calculation of the sheet resistance ($R_{sheet}$). The details for the measurement procedure presented in ESI (fig.S1,S2; eq.S2-S7 in ESI).

Measured data and calculated parameters presented in **Tab.1**. Mainly CR-ITO and SLG/ITO differed in concentration of the charge carriers (**n**) and mobility, which impacted conductivity ($\sigma$). While SLG ITO the n was at $\sim 10^{20}$ cm$^{-3}$, the value for CG/ITO reduced to $5 \times 10^{19}$ cm$^{-3}$. Hall mobility values for SLG and CG ITO were 44 cm$^2$V$^{-1}$s$^{-1}$ and 66 cm$^2$V$^{-1}$s$^{-1}$, respectively. While CG/ITO does demonstrate slightly higher mobility, the difference is not enough to offset the substantial decrease in carrier concentration and reduction of conductivity.

Table 1. Electric transport properties of the ITO coatings deposited on SLG and CG, respectively

| Sample | Specific conductivity ($\sigma$), S/cm | Hall coefficient (R), m$^3$/Cl | Hall mobility, cm$^2$/(V·s) | Concentration of the charge carriers (n), cm$^3$ | Sheet resistance ($R_{sheet}$), Ohm/sq |
|---|---|---|---|---|---|
| SLG/ITO | $8.9 \pm 0.2 \cdot 10^2$ | $5.9 \pm 0.3 \cdot 10^{-8}$ | $44.6 \pm 1.8$ | $1.12 \pm 0.01 \cdot 10^{20}$ | $86.7 \pm 1.2$ |
| CG/ITO | $5.61 \pm 0.16 \cdot 10^2$ | $1.39 \pm 0.03 \cdot 10^{-7}$ | $65.8 \pm 1.5$ | $5.34 \pm 0.02 \cdot 10^{19}$ | $137.0 \pm 2.6$ |

*Hall measurements were performed in 0.22T magnetic field for 130 nm thick ITO films*

Hall mobility in ITO films is distinctly sensitive to microstructure, crystallinity, carrier concentration, grain boundary scattering, and point defects (e.g., oxygen vacancies, tin interstitials). As reported, for nanocrystalline ITO, Hall mobility is typically found between 10 and 80 cm$^2$V$^{-1}$s$^{-1}$[24,25], with higher values generally occurring at lower carrier concentration. In general, carrier concentration and mobility in ITO have an inverse correlation due to ionized dopants (Sn$^{4+}$ replacing In$^{3+}$) or defects (vacancies) scattering. When n decreases, µ can increase—but only up to a point before other scattering mechanisms become dominant. Ionized impurities form a screened Coulomb potential that scatter conduction electrons. As carrier concentration increases (either by more Sn or more oxygen vacancies), there are more scattering centers, which reduces the mobility. This is consistent with the observed data. The reduced charge carrier concentration typically improves the transmittance of ITO originated from decreased free carrier absorption. In our case, the CG/ITO (n~$10^{19}$ cm$^3$) has obviously affected transmittance compared to SLG/ITO (n~$10^{20}$ cm$^3$). So, lower transparency in the low-n film suggests a contribution of factors such as microstructure and/or imperfect stoichiometry, relied on tin and oxygen content. C-AFM mapping shown on **Fig. S3** (**ESI**) didn't reveal significant inhomogeneities of the surface conductivity for both samples and the observed features are associated with the topography induced changes in the contact area.

We correlated the data with the energy-dispersive X-ray spectroscopy (**EDX**) measurements to gather insight into changes of the obtained electrical properties. The oxygen content has been considered affecting the charge transport of ITO films, as well as Sn–as acceptor agent. As reported by *Buchanan et al.*[26], the carrier density is ruled by the oxygen vacancies. In the presence of such defects, the ITO has a reduced concentration of the charge carriers, and the resistivity increases. According to our data (**tab.2** with atomic concentration values), the CG/ITO exhibited an atomic concentration of oxygen at 49.9%, while reference deposited on soda lime had 48.7%. Thus, the key-difference in electric properties observed for coating on cerium glass originated from the reduced concentration of oxygen and tin. This points out the specifics of the surface properties for the growth of nanocrystalline ITO films.

Table 2. Atomic concentration of ITO on SLG and CG substrates extracted from EDX measurements

| Sample | Atomic concentration % | | | | | | | | | In/Sn |
|---|---|---|---|---|---|---|---|---|---|---|
| | O | Na | Si | In | Sn | Ce | Ca | K | Mg | |
| SLG | 48.72 | 6.75 | 31.44 | 6.25 | 0.83 | 0 | 3.61 | 0 | 1.76 | **7.53** |
| CG | 49.96 | 5.14 | 33.09 | 7.23 | 0.88 | 0.35 | 0 | 3.35 | 0 | **8.22** |

It is well established that the indium-to-tin atomic ratio of 9:1[27–30]. However, compositional analysis of our samples revealed a significant deviation from this stoichiometry. In the SLG/ITO sample, the measured In/Sn ratio was 7.53, while in the CG/ITO sample it was 8.22. This indicates an excess of tin incorporated into the ITO structure, which is known to adversely affect both the electrical transport properties and the optical transparency of the films. We attribute this excess tin content and its variation between substrates to differences arising during the deposition process. In ion-beam sputtering, the ion flux interacts with the growing film through both elastic and inelastic collisions with the substrate surface. A fraction of the incoming ions doesn't contribute to film growth[31,32] but induces preferential sputtering of indium atoms from the growing film. As a result, the deposited ITO films exhibit indium deficiency and a relative enrichment of tin. Furthermore, the observed difference in In/Sn ratio between SLG/ITO and CG/ITO samples is likely associated with the variation in the interaction dynamics at the film-substrate interface. The bonding strength and energy dissipation mechanisms differ between soda-lime glass (SLG) and cerium-doped glass (CG), affecting the extent of indium re-sputtering and incorporation efficiency during film growth. This substrate-dependent effect leads to a more pronounced indium loss in the SLG/ITO samples, resulting in a lower In/Sn ratio compared to CG/ITO.

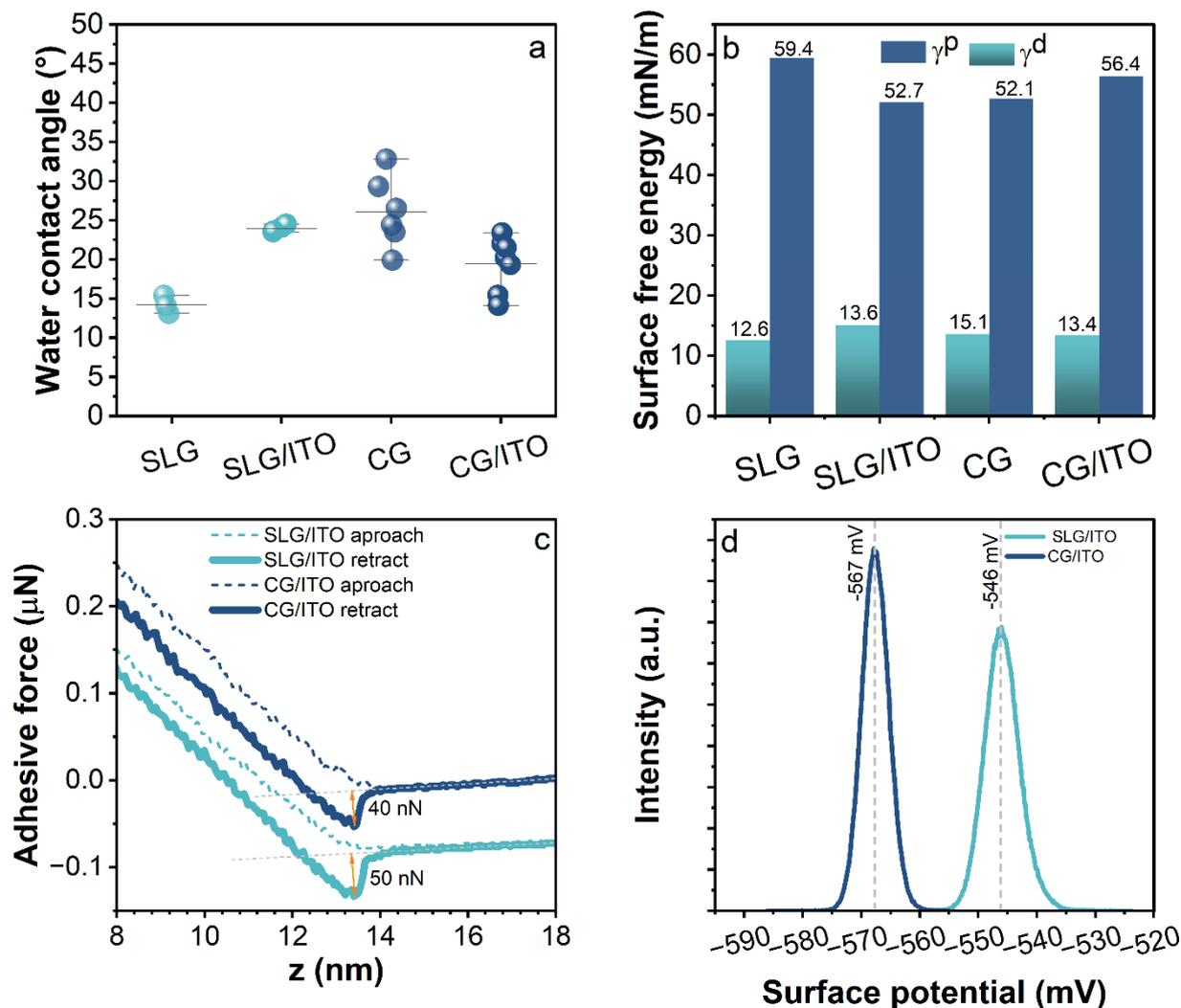

Figure 3 – Water contact angle on SLG and CG glass substrates and deposited ITO (a), surface free energy glasses and ITO film, adhesion force curves of ITO films (c) and surface potential distribution on ITO surface (d)

The surface properties of ITO films deposited on SLG and CG substrates were comprehensively investigated using water contact angle measurements, surface free energy analysis, AFM-based adhesion force measurements, and Kelvin probe surface potential mapping, as shown in **fig.3**. The bare SLG and CG substrates exhibit low contact angles (**fig.3(a)**) of 15° and 25°, respectively, indicating hydrophilic behavior. After ITO deposition, the contact angle on SLG increases slightly to ~25°, suggesting a moderate decrease in surface wettability. In contrast, for the CG/ITO sample, the contact angle slightly decreases to ~20°, indicating an increase in surface free energy after ITO deposition, that was quantified using the Owens-Wendt method[33] with H$_2$O, DMSO, and DMF as probe liquids. The calculated polar ($\gamma^p$) and dispersive ($\gamma^d$) components are summarized in **Fig. 3(b)**. All samples exhibit high total surface energy, predominantly governed by the polar component. Deposition of ITO on SLG reduces the total surface energy by 5.7 mN/m. Conversely, the crystallization of a more textured and oxygen-deficient ITO layer on CG results in a slight increase in total surface energy by 2.6 mN/m, primarily due to an increase in the polar component. This reflects substrate-induced differences in surface polarity following ITO deposition. Adhesion force measurements derived from AFM force–distance curves are shown in **Fig. 3(c)**. The figure shows typical curves picked from the measured 100 curves. The retraction curves reveal adhesive forces of approximately 40±20 nN for SLG/ITO and

50±20 nN for CG/ITO. Because of the tip curvature radius comparable with grain sizes on the samples surface the measured adhesion values were in the relatively wide range. This difference in tip–surface interaction is likely related to the interaction of the surface with adsorbed water layers from the ambient environment. The slightly higher adhesion force in CG/ITO correlates with the increased surface free energy.

Kelvin probe surface potential measurements (**Fig. 3(d)**) show a clear difference between the two samples. The CG/ITO surface exhibits a more negative surface potential (−567±10 mV) compared to SLG/ITO (−546 mV±10), suggesting a Fermi level pinning[34–37] induced by defect and coating/substrate strain effects. Lower work function values were also observed due to the excess tin, like in the SLG/ITO case[38]. Surface potential maps presented in **Fig. S4 (ESI)** revealed a uniform potential distribution for both samples, which is consistent with the result of the C-AFM mapping.

In conclusion, the results of this systematic study provide important insights for reactive producing ITO coatings on Ce-doped glass that are potentially applicable to perovskite solar cells in space. Ce-glass could be used as a relevant lightweight substrate with an electrode. The Hall mobility of the charge carriers in CG/ITO reached 65.8 $cm^2V^{-1}s^{-1}$, while the coating had preferential (222) orientation for nanocrystallites, which is typical for room-temperature deposition processing[39–41]. However, compared with standard soda-lime glass (SLG), the Ce-doped analogue alters the growth process of the ITO layer with several critical features. The presence of cerium in the glass structure modifies energy-dissipation pathways during deposition, resulting in indium re-sputtering effects, changes in the In/Sn ratio, and presence of the compressive lattice strain at film–substrate interface. Structural and stoichiometric changes in CG/ITO relative to the SLG reference also show higher oxygen and tin content, which leads to reduced charge-carrier concentration and lower film conductivity. This implies that, for Ce-doped glass, it is advisable to use targets with a reduced tin content compared with the standard In:Sn ratio of 9:1. Introducing oxygen into the working gas (Ar) could potentially increase the optical transmittance to values >80%, as desired for the front transparent electrode of solar cells. Nevertheless, strong intrinsic absorption in the UV and short-wavelength region, even for 100-um glass, indicates the need for antireflection coatings (e.g., $MgF_2$, $Si_3N_4$) to mitigate these parasitic optical losses. Additionally, the emergence of island-like inhomogeneities in the sputtered ITO film on CG indicates the need for surface pretreatment, which may include high-power plasma processing to reduce the surface energy.

**SUPPLEMENTARY MATERIAL**

The supplementary material includes description of the materials section (materials, characterization), AFM images, c-AFM images and KPFM maps of perovskite surface. The data that supports the findings of this study are available within the article and its supplementary material.

The authors gratefully acknowledge the financial support from the Russian Science Foundation with project № 24-62-00022.

**Author Contributions**

**D.D.G.** - gathering, methods, analysis, writing original manuscript.

**L.L.O.** - gathering, methods, analysis, writing original manuscript.

**D.Yu.D.** - gathering, methods,

**P.A.A.** data gathering, methods, analysis, writing original manuscript.

**I.R.S.** – data gathering, methods.

**A.R.T.** – data gathering, methods.

**M.S.D** – data gathering, methods, analysis, writing original manuscript.

**V.K.** – data gathering, methods.

**V.I.** – data gathering, methods.

**D.S.S.** - coordinated the research activity, analysis, writing original manuscript.

**E.I.T. –** funding acquisition, supervision, writing original manuscript.

The manuscript was written with contributions from all the authors. All the authors approved the final version of the manuscript.

Electronic supplementary information (ESI) for the paper:

# Specifics of ITO properties deposited on cerium-doped glass for space-grade solar cells

Danil D. Gren[1], Lev O. Luchnikov[1], Dmitri Yu. Dorofeev[1], Prokhor A. Alekseev[2,3], Ildar R. Sayarov[4], Alexey R. Tameev[4], Mikhail S. Dunaevskiy[2,3], Vladislav Kalinichenko[2], Vladimir Ivanov[2], Danila S. Saranin[1,2]* and Eugene I. Terukov[2]

[1]National University of Science and Technology MISIS Moscow, Russia

[2]ITMO University, 197101, St. Petersburg, Russia

[3]Ioffe Institute, 194021, St. Petersburg, Russia

[4]Frumkin Institute of Physical Chemistry and Electrochemistry, Russian Academy of Sciences,

**Corresponding author\*:**

Dr. Danila S. Saranin saranin.ds@misis.ru

**Experimental section:**

*Materials*

Ion-beam sputtered ITO films were fabricated with use of In:Sn (90%:10%) target (99.99 purity%) purchased from LLC Girmet (Russia). Soda lime glass was purchased from LLC Khimmed with a thickness of 1.1 mm. Ce-doped glass of 100 micron thickness was purchased from LLC SMK-Steklo Company (Russia). The $CeO_2$ concentration in the glass was 1%. Quartz glass of KU-1 grade (2 mm) was purchased from LLC Elektrosteklo Company (Russia).

*ITO deposition parameters*

The ion-beam sputtering machine was NIKA-13 from Beams and Plasmas Company (Russia). We set an ion beam source positioned in the vacuum chamber at an 8° angle relative to the horizontal plane. The alloyed In-Sn target of rectangular shape is placed on the water-cooled pedestal with a tilt angle = 13°. The substrates with samples are mounted horizontally on the carousel with lateral movement. Uniformity of the fabricated coating was achieved by moving the substrate holder over the deposition window.

The following parameters were used for the deposition of ITO: base pressure - $2\times10^{-2}$ Pa; working pressure – 0.10 Pa; working gas – Ar (99.998%); the source voltage - 3000 V; current in the ion beam source – 130 mA; the distance between center of the target and the substrate holder – 120 mm. Prior to the process, the target was etched with an ion beam for 10 minutes to remove the oxidized surface layer, organics, and water impurities.

*Characterization*

Surface roughness and films thicknesses were measured with KLA-Telencor stylus profilometer. X-Ray diffraction **(XRD)** of perovskite layers was investigated with diffractometer Tongda TDM-10 using CuKα as a source with wavelength 1.5409 Å under 30 kV voltage and a current of 20 mA. absorbance spectra **(ABS)** of ITO films were carried out via SE2030-010-DUVN spectrophotometer with a wavelength range of 300–1100 nm. Atomic force microscopy **(AFM)** and Kelvin probe force microscopy **(KPFM)** measurements were performed in room ambient conditions using Ntegra (NT-MDT) microscope. NSG10/Pt (Tipsnano) probes were used with tip curvature radius 30 nm.

## Measurement and Calculation Methodology

### *Tauc plots analysis*

The Tauc plots calculation was realized using equation:

$$(\alpha h\nu)^2 = A(h\nu - E_g) \quad (S1)$$

where α is absorption coefficient being a function of wavelength $\alpha(\lambda)$, $h$ is Planck constant, $E_g$ is an optical band gap of a semiconductor, $\nu$ is frequency, $A$ is proportionality constant, and $n$ is Tauc exponent.

### *Four-probe measurements*

We measured the series resistance using four-point probe station "VIK-UES" (MEDNM, Russia). Four in-line tungsten carbide probes with a spacing of 0.75 mm are used, the applied current was 100 uA. Each sample was measured in the center four times, to obtain an average series resistance of the film and estimate the spread. Systematic error in the measurement system does not exceed 2%.

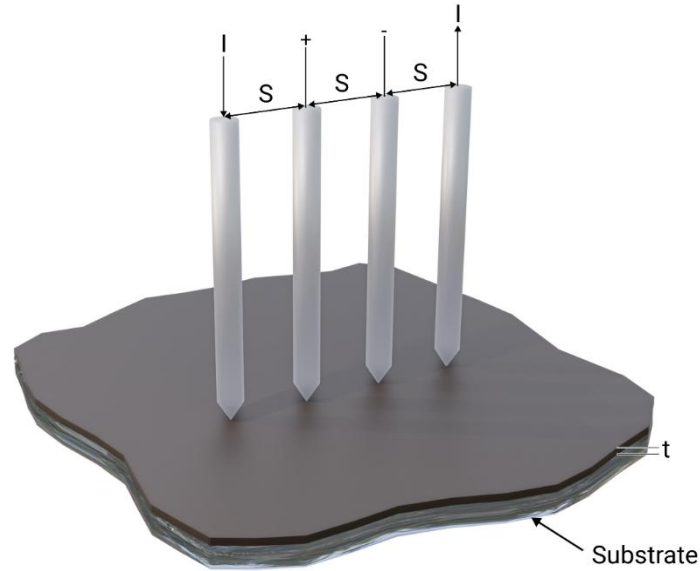

Figure S1 – scheme of four-probe measurement

### *Hall measurements*

The mobility of charge carriers in ITO films was measured based on the Hall effect in a weak magnetic field (eq.S2).

$$E_H = -R_H[j, B] \quad (S2)$$

where $R_H$ is the Hall constant determined by the properties of the sample, $E_H$ is the Hall field

Using the relationship between the intensity $E_H$ and the potential difference $U_H$ we obtain eq.S3:

$$R_H = \frac{U_H d}{IB} \quad (S3)$$

In the general case where the relaxation time of the electrons depends on their velocity, it is necessary to consider the scattering mechanism and write down the expression for the Hall coefficient taking into account the multiplier r, called the Hall factor, which varies from 1 to 2 (eq.S4):

$$R_H = -\frac{r}{en} \quad (S4)$$

Then the Hall mobility and concentration of charge carriers can be calculated by the formulas(eq.S5):

$$\mu_H = |R|\sigma \quad (S5)$$

The mobility of the charge carriers can be obtained from formula (S3) and (S5):

$$\mu = \frac{\mu_H}{r} \quad (S6)$$

In this work, a Keithley 2400 source-meter was used as a direct current source, and the Hall voltage was recorded using a Keithley 236 voltmeter. The equivalent electric circuit is shown in Fig. S2. Probes placed on the vertices of a square with a side of 3.4 mm were used as contacts. Measurements were performed in the absence of a magnetic field and at a constant magnetic field B=280 mTl.

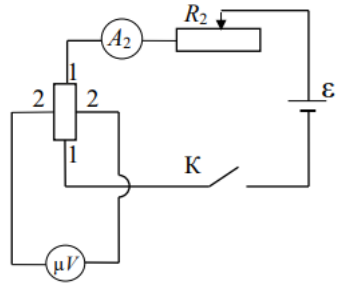

Figure S2 - The equivalent electric circuit

The Hall voltage was calculated by formula (S7):

$$U_H = \frac{U^N - U^S}{2} \quad (S7)$$

Where $U^N$ and $U^S$ are voltages on the voltmeter at opposite directions of the magnetic field.

Then the average values of the Hall constant and the Hall mobility were calculated using formulas (S3), (S5). The coefficient r depends on the scattering mechanism. At low temperatures the impurity ions play the main role in the scattering and r = 1.92 should be assumed. Since the measurements were carried out at room temperature, we will assume that the scattering was mainly carried out on acoustic phonons. Then we take r = 1.18

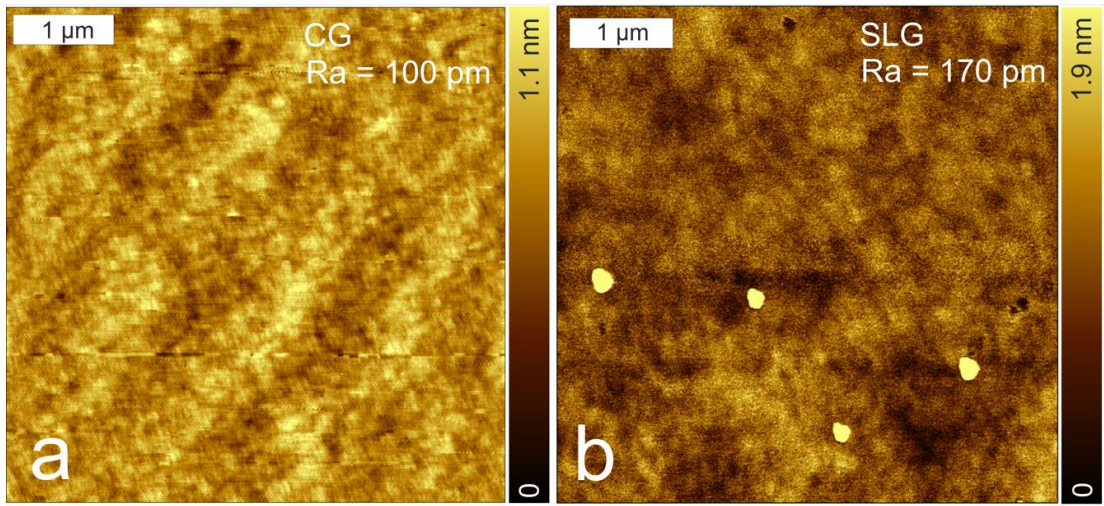

Figure S3 – Surface morphology of CG (a) and SLG (b)

Atomic force microscopy (AFM) was used to examine the surface morphology of glass substrates, specifically cerium-doped glass (CG) and soda-lime glass (SLG), as shown in Fig. 1a and 1b. The CG samples exhibited a uniform and smooth surface with an average roughness (Ra) of approximately 100 pm. In contrast, SLG samples also displayed a generally homogeneous morphology, but with clearly distinguishable nanoscale crystalline inclusions, resulting in an increased average roughness of 170 pm. AFM imaging was performed in semi-contact (tapping) mode using a AIST Smart SPM 1000 system (NT-MDT, Russia). A silicon cantilever of type NGS30 (NT-MDT, Russia) was employed, with an oscillation amplitude of 40 nm and a cantilever length of 125 nm.

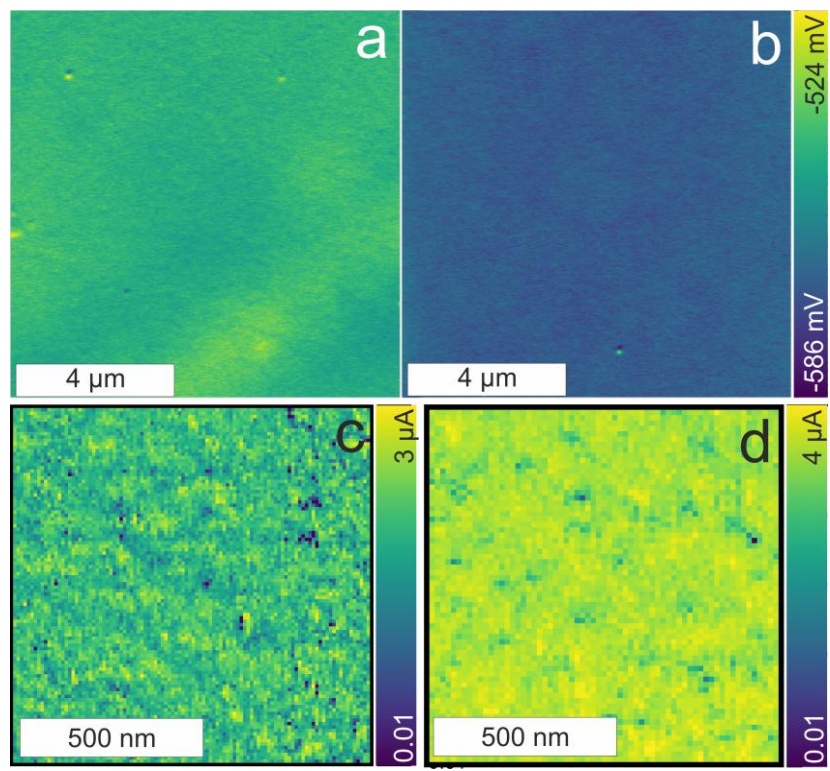

Figure S4 – maps of surface potential of ITO on SLG (a) and CG(b), surface current maps of SLG/ITO (c) and CG/ITO (d) samples